\documentclass[a4paper,11pt]{article}

\usepackage{rotating}
\usepackage{graphics}
\usepackage{latexsym}
\usepackage[dvips]{epsfig}
\usepackage{amssymb}
\usepackage{graphicx}

\usepackage{url}

\def\boxit#1{\vbox{\hrule\hbox{\vrule\kern4pt
  \vbox{\kern1pt#1\kern1pt}
\kern2pt\vrule}\hrule}}

\def\qed{\rule{1.5mm}{3mm}}

\newcommand\nc{\newcommand}

\newtheorem{theorem}{\bfseries Theorem}
\newtheorem{lemma}[theorem]{Lemma}

\newtheorem{defn}[theorem]{Definition}

\nc{\crl}[2]{\begin{corollary}\label{crl:#1} #2 \end{corollary}}
\nc{\dfn}[2]{\begin{defn}\label{def:#1} #2 \end{defn}}
\nc{\lem}[2]{\begin{lemma}\label{lem:#1} #2 \end{lemma}}
\nc{\prp}[2]{\begin{proposition}\label{prp:#1} #2
\end{proposition}}
\nc{\thm}[2]{\begin{theorem}\label{thm:#1} #2\end{theorem}}
\nc{\fac}[2]{\begin{lemma}\label{fact:#1} #2 \end{lemma}}

\nc{\eqn}[2]{\begin{eqnarray}\label{eqn:#1} #2 \end{eqnarray}}

\nc{\fig}[4]{\begin{figure}[h]
\begin{center}
\includegraphics[width=#2\textwidth]{#4}
\end{center}
\caption{#3}\label{fig:#1}
\end{figure}}

\nc{\tbl}[3]{\begin{table}[hbt] #3 \caption{#2} \label{tab:#1}
\end{table}}

\nc{\refc}[1]{Corollary~\ref{crl:#1}}
\nc{\refd}[1]{Definition~\ref{def:#1}}
\nc{\reff}[1]{Fig.~\ref{fig:#1}}
\nc{\refl}[1]{Lemma~\ref{lem:#1}}
\nc{\refp}[1]{Proposition~\ref{prp:#1}}
\nc{\reft}[1]{Theorem~\ref{thm:#1}} \nc{\refe}[1]{(\ref{eqn:#1})}
\nc{\reftb}[1]{Table~\ref{tab:#1}}

\nc{\reffc}[1]{Fact~\ref{fact:#1}}

\nc{\pf}[1]{ \noindent \emph{Proof.} #1
 \hfill \qed\par}

\renewcommand{\title}[1]{\vspace{\fill}
\eject\addtolength{\baselineskip}{4pt}
{\bfseries\Large #1}\\[3mm]\addtolength{\baselineskip}{-4pt}}
\renewcommand{\author}[3]{\parbox[t]{75mm}
{\begin{center}{\scshape #1}\\[3mm] #2\\
 {\ttfamily #3} \end{center}}}

\long\def\invis#1{}
\begin{document}

\begin{center}
\title{Some Reduction Operations to Pairwise Compatibility Graphs}

\author{Mingyu Xiao}
 {School of Computer Science and Engineering,
University of Electronic Science and Technology of China, China,
}{myxiao@gmail.com}
\author{Hiroshi Nagamochi}{
Department of Applied Mathematics and Physics,
  Graduate School of Informatics, Kyoto University, Japan,
}{nag@amp.i.kyoto-u.ac.jp}
%
\end{center}

\begin{abstract}
A graph $G=(V,E)$ with a vertex set $V$ and an edge set $E$
 is called a  pairwise compatibility graph (PCG, for short)
if there are  a tree $T$ whose leaf set is $V$,
a non-negative edge weight $w$ in $T$,
and two non-negative reals $d_{\min}\leq d_{\max}$
such that $G$ has an edge $uv\in E$ if and only if
the distance between $u$ and $v$ in the weighted tree $(T,w)$ is
in the interval $[d_{\min}, d_{\max}]$.
PCG is a new graph class motivated from bioinformatics.
In this paper, we give some
 necessary and sufficient conditions for PCG based on cut-vertices and twins,
 which provide reductions among  PCGs.
Our results imply that complete $k$-partite graph,
cactus, and some other graph classes are subsets of PCG.

\vspace{5mm}\noindent {\bf Key words.}
Graph Algorithms; Pairwise Compatibility Graph; Graph Theory; Reduction \ \
\end{abstract}

\section{Introduction}
\begin{figure}[htbp] \begin{center}
 \includegraphics[scale=1.0]{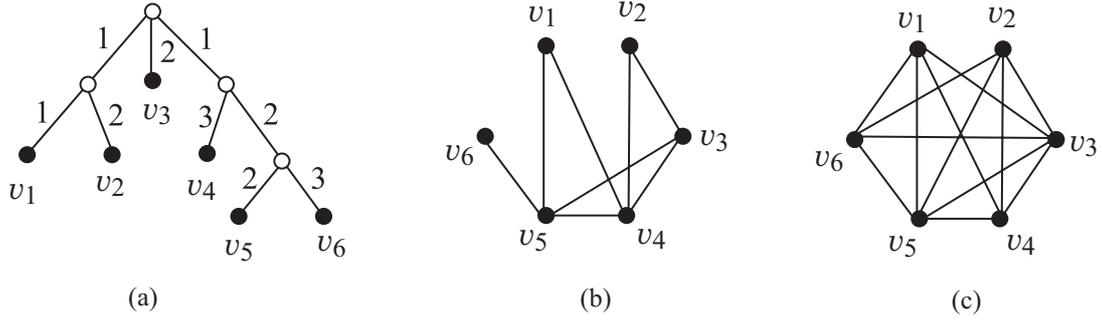} \end{center}
 \caption{(a) An edge-weighted tree $(T,w)$,
 (b) The PCG obtained from $(T,w)$ and $(d_{\min}, d_{\max})=(5,7)$, and
 (c) The PCG obtained from $(T,w)$ and $(d_{\min}, d_{\max})=(4,8)$.} \label{f1:example1}
 \end{figure}

An unweighted simple undirected  graph $G=(V,E)$
with a vertex set $V$ and an edge set $E$ is called
a \emph{pairwise compatibility graph} (PCG, for short)
 if there exist a tree $T$
 with edges weighted by non-negative reals and two non-negative real numbers
$d_{\min}$ and $d_{\max}$ such that: the leaf set of $T$ is $V$,
and  two vertices $u,v\in V$ are adjacent in $G$ if and only if the distance between
$u$ and $v$  in   $T$ is at least $d_{\min}$ and at most $d_{\max}$.
The tree $T$ is also called a \emph{pairwise compatibility tree} (PCT, for short) of the graph $G$.
The same tree $T$ can be a PCT of more than one PCG.
Figure~\ref{f1:example1} shows an edge-weighted tree $(T,w)$
 and two PCGs for $(T,w)$ with $(d_{\min}, d_{\max})=(5,7)$ and $(d_{\min}, d_{\max})=(4,8)$ respectively.
The concept of PCG was first introduced by Kearney et al.~\cite{phtree} to model evolutionary relationships among a set of organisms in bioinformatics.
However, it is a challenging problem to construct a pairwise compatibility tree for a given graph.
Recognition and characterization of PCGs became interesting problems in graph theory recently.

Not every graph is a PCG. Yanhaona, Bayzid and Rahman~\cite{DiscoveringPCG}
constructed the first non-PCG, which is a bipartite graph with 15 vertices.
Later, an example with 8 vertices was found in~\cite{noPCG}.
This is the smallest non-PCG, since
it has been checked that all graphs with at most seven vertices are PCGs~\cite{CFS2013}.
Currently no polynomial-time algorithm is known to the problem of testing
 whether a given graph is a  PCG or not.
 It is widely believed that  recognizing PCGs  is NP-hard~\cite{survey,noPCG}.

In the literature, there are several contributions to recognizing some subclasses of PCG.
It is not difficult  to see that every tree is a PCG~\cite{conditions}.
Every cycle with at most one chord
 has also been shown to be a PCG~\cite{Y-PCG}.
Other subclasses of graphs currently known as PCGs  are
power graphs of trees~\cite{DiscoveringPCG},
threshold graphs~\cite{mLPG},
triangle-free outerplanar 3-graphs~\cite{triang-free},
a special
subclasses of split matrogenic graphs~\cite{splitgraph},
Dilworth 2 graphs~\cite{dilworth2,dilworthk},
the complement of a forest~\cite{conditions},
the complement of a cycle~\cite{exploring} and so on.
Some conditions for a graph not being a PCG have also been developed~\cite{noPCG,conditions,DiscoveringPCG,revisited}.
However, there is still few known method for generating PCGs and PCTs with complicated structures.

In this paper, we will give more
 necessary and sufficient conditions for PCG.
 The first one is related to cut-vertices, where we show that
 a graph is a PCG if and only if each biconnected component
 of it is a PCG.
  The second one is about a pair of vertices  with the same neighbors, called  ``twins.''
   We will show some conditions under which we can add a copy $v'$ of a vertex $v$ into a PCG
   so that $v'$ and $v$ form twins
    to get another PCG.
    One of our results answers an open problem on ``true twins'' \cite{exploring}.
 These properties provide simple reductions rules,
 by which we can reduce some graph into a smaller graph to check if it is a PCG
 and   find more subclasses of PCGs as well as non-PCGs with an arbitrary large size.
  For examples, our results imply that complete $k$-partite graphs,
 cacti, and some other graphs are subclasses of PCG.

\section{Preliminary}\label{sec:preliminary}

Let a graph $G=(V,E)$ stand for an unweighted simple undirected
graph with sets $V$ and $E$ of vertices and edges,
respectively.
An edge with end-vertices $u$ and $v$ is denoted by $uv$.
For a graph $G$, let $V(G)$ and $E(G)$ denote the sets of vertices and edges
in $G$, respectively, and let $N_G(v)$ be the set of neighbors of a vertex $v$ in $G$ and let $N_G[v]=N_G(v)\cup\{v\}$.
Two vertices $u$ and $v$ in a graph $G$
are \emph{true twins} (resp., \emph{false twins}) if $N_G[v]=N_G[u]$ (resp., $N_G(v)=N_G(u)$).
For a subset $X\subseteq V(G)$, let $G-X$ denote
the graph obtained from $G$ by removing vertices in $X$
together with all edges incident to vertices in $X$,
where $G-\{v\}$ for a vertex $v$ may be written as $G-v$.
Let $G[X]$ denote the graph induced by a subset $X\subseteq V(G)$,
i.e., $G[X]= G -(V(G)\setminus X)$.

A vertex is called a \emph{cut-vertex}  if   deleting it increases the number of connected component of the graph.
A graph is \emph{biconnected} if it has no cut-vertex.
Note that a graph consisting of a single edge is biconnected.
A \emph{biconnected component} in a graph is a maximal biconnected subgraph.
A \emph{cactus} is a connected graph in which any two simple cycles have at most one vertex in common.
Note that each biconnected component of a cactus is either a cycle or an edge.
A graph is called a \emph{complete $k$-partite graph}
if the vertex set can be partitioned into $k$ disjoint non-empty vertex subsets such that
no two vertices in the same subset are adjacent whereas
 any two vertices from different subsets
are adjacent.
A complete $k$-partite graph with
$k$ subsets $V_1,V_2,\ldots,V_k$ with $|V_i|=s_i$ is denoted by $\mathrm{K}_{s_1,s_2,\dots, s_k}$.

Let $T$ be a tree.
A vertex in a tree
is called an {\em inner vertex} if it is incident to at least two edges
and is called a {\em leaf} otherwise.
Let $L(T)$ denote the set of leaves in the tree $T$.
An edge incident to a leaf in $T$ is called a {\em leaf edge} of $T$.
For a subset $X\subseteq V(T)$ of vertices,
let $T\langle X\rangle$ denote a minimal subtree
of $T$
 subject to the condition that
any two vertices $u,v\in X$ remain connected in  $T\langle X\rangle$.
Note that for a given subset $X$, the minimal subtree is unique.

An edge-weighted graph $(G,w)$ is defined to be a pair of
a graph $G$ and a non-negative weight function $w:E(G) \to \Re_+$.
 For a subgraph $G'$ of $G$, let $w(G')$ denote
 the sum $\sum_{e\in E(G')}w(e)$ of edge weights in $G'$.

Let $(T,w)$ be an edge-weighted tree.
For two vertices $u,v\in V(T)$,
the {\em distance} $\mathrm{d}_{T,w}(u,v)$ between them is defined to be
 $w(T\langle \{u,v\}\rangle)$, i.e.,
  the sum of weights of edges in the path
between $u$ and $v$ in $T$.

For a tuple $(T,w,d_{\min}, d_{\max})$ of an edge-weighted tree $(T,w)$ and
two  non-negative reals $d_{\min}$ and $d_{\max}$,
define $G(T,w,d_{\min}, d_{\max})$ to be the simple graph
$(L(T), E)$ such that, for any two distinct vertices $u,v\in L(T)$,
$uv\in E$ if and only if $d_{\min} \leq \mathrm{d}_{T,w}(u,v)\leq d_{\max}$.
We define $E$ to be an empty set if $|V(T)|=1$.
Note that $G(T,w,d_{\min}, d_{\max})$ is not necessarily connected.
For a subset $X\subseteq V(T)$,
 let $w_X: E(T\langle X\rangle) \to \Re_+$ be
 a function such that $w_X(e)=w(e)$, $e\in E(T\langle X\rangle)$,
 where we regard $w_X$ as null if $|X|\leq 1$.

A graph $G$  is called a {\em pairwise compatibility graph} (PCG, for short)
if there exists a tuple $(T,w,d_{\min}, d_{\max})$ such that
$G$ is isomorphic to the graph $G(T,w,d_{\min}, d_{\max})$,
where we call such a  tuple a   {\em pairwise compatibility representation} (PCR, for short) of $G$,
and call a tree $T$ in a PCR  of $G$ a {\em pairwise compatibility tree}
 (PCT, for short) of $G$.
We call $d_{\min}$ and $d_{\max}$ the {\em lower and upper bounds} of a PCR.

\section{Some Structures on PCR}\label{sec:basic}

We start to review the following property, which has been frequently used in literature.
The correctness of it  immediately follows from the definition of PCG.

\begin{lemma}\label{le:basic}
Let $(T,w,d_{\min}, d_{\max})$ be  a PCR  of a graph $G$.
For any subset $X\subseteq V(G)$,
the tuple
  $(T\langle X\rangle, w_X, d_{\min}, d_{\max})$   is a PCR of the induced graph $G[X]$.
\end{lemma}

A PCR $(T,w,d_{\min}, d_{\max})$ of a PCG is called \emph{non-singular} if
$T$ contains at least three vertices,
$0<d_{\min}< d_{\max}$, and $w(e)>0$ holds for all edges $e\in E(T)$.

\begin{lemma}\label{le:1}
Let $G$ be a PCG with at least two vertices.
Then $G$ admits a non-singular PCR.
Given a PCR of  $G$,
a non-singular PCR of   $G$ can be constructed in linear time.
\end{lemma}
\noindent {\bf Proof.}
Let $G$ be a PCG with $|V(G)|\geq 2$
and $(T,w,d_{\min}, d_{\max})$ be an arbitrary PCR of $G$.
 We will construct a non-singular PCR  of $G$ by four steps below.

First, if there is a non-leaf edge $e$ such that $w(e)=0$, we
can shrink it by identifying the two end-vertices of it.
The resulting graph is still a tree, a leaf in the original is still a leaf in this tree,
and the distance between any two vertices in the tree remains unchanged.
So the new tree is still
a PCT of the graph $G$.
Now we assume that the edge weight of any non-leaf edge in the tree is positive.

By assumption of $|V(G)|\geq 2$, it holds that $|V(T)|\geq |L(T)|=|V(G)|\geq 2$.
Next if $|V(T)|=2$,
 then we subdivide the unique edge $uv$ in $T$ with a new inner vertex $v^*$
 so that $w'(uv^*)+w'(v^*v)=w(uv)$ in the new tree $T'$ obtained by subdividing the edge $uv$.
 It is easy to see that the new tuple $(T',w',d_{\min}, d_{\max})$ is still a PCR of $G$,
 and $|V(T')|\geq 3$.
 In the following we assume that a PCT  has at least three vertices.

 In a PCR $(T,w,d_{\min}, d_{\max})$ with $|V(T)|\geq 3$,
each path between two leaves contains exactly two leaf edges.
As for the third step, if $w(e)=0$ for some leaf edge $e\in E(T)$ or $d_{\min}=0$,
then we can change all leaf edge weights and $d_{\min}$
 positive if necessary, by increasing the weight of each leaf edge by a positive real $\delta>0$
 and increasing each of  $d_{\min}$ and $d_{\max}$ by $2\delta$.
 The resulting tuple is a PCR of the same graph $G$.
 Now all of edge weights, $d_{\min}$ and $d_{\max}$ are positive.

 Finally, if the lower and upper bounds are same, i.e., $d_{\min}=d_{\max}$, then
 we augment the upper bound $d_{\max}$ to $d'_{\max}:=d_{\max}+ \varepsilon$ by choosing
 a sufficiently small  positive real $\varepsilon$
 so that every two  leaves $u$ and $v$ in $T$ such that $\mathrm{d}_{T,w}(u,v)> d_{\max}$
 still satisfies $\mathrm{d}_{T,w}(u,v)> d_{\max}+\varepsilon~(=d'_{\max})$.
 Obviously the resulting tuple with $d'_{\max}$ is a PCR of the same graph $G$
 and satisfies
 $d_{\min}\neq d'_{\max}$.

 After executing the above four steps, we can get a non-singular PCR of the graph $G$.
 Furthermore, all the four steps can be done in linear time.
\qed
\bigskip

A PCR $(T,w,d_{\min}, d_{\max})$ of a PCG is called \emph{normalized} if
$0< d_{\min}< 1$, $ d_{\max}=1$, $w(e)>0$ holds for all edges $e\in E(T)$, and
 $w(e)>1/4$ holds for all leaf edges $e$ in $T$.
We have the following lemmas.

\begin{lemma}\label{le:2}
Let $G$ be a PCG with at least two vertices.
Then there is a positive constant $c_G$ with $1/2< c_G<1$ such that
for any real $\alpha$ with $ c_G<\alpha<1$,
 $G$ admits a normalized PCR with $(d_{min},d_{max})=(\alpha,1)$.
Given a PCR of $G$,
such a normalized PCR of  $G$ can be constructed in linear time.
\end{lemma}
\noindent {\bf Proof.}
By Lemma~\ref{le:1}, we know that   a non-singular PCR $(T,w,d_{\min}, d_{\max})$ of $G$
can be constructed in linear time.
For $c_G=\frac{d_{\min}+d_{\max}}{d_{\max}+ d_{\max}}$, where $1/2< c_G<1$,
let $\alpha$ be any real such that  $c_G<   \alpha<1$.
To prove the lemma, it suffices to show that a normalized PCR $(T,w',\alpha, 1)$ can be constructed in linear time.

Let $\delta$ be the positive real such that
  $\frac{d_{\min}+ \delta}{d_{\max}+ \delta}=\alpha$, where  $\delta> d_{\max}$ holds.
  We increase the weight of each leaf edge  in $T$
  by   $\delta/2$,
  which increases the weight of each path between
  two leaves in $T$ by $\delta$. We scale the weight in the tuple so that
  the lower and upper bounds become $\alpha$ and 1; i.e., we
  divide by $d_{\max}+ \delta$  the weight of each edge in $T$ and each of $d_{\min}+ \delta$
  and     $d_{\max}+ \delta$.
  This results in a tuple  $(T, w', \alpha, 1)$ of $G$
    such that $w'(e)\geq (\delta/2)/(d_{\max}+ \delta)>1/4$ for each  leaf edge  $e$  in $T$.
\qed
\bigskip


Most of our arguments are based on normalized PCR,
since it will be helpful for us to simplify some proofs.

\section{Properties on Induced Subgraphs of PCGs}\label{sec:rule}

In this section, we derive some sufficient conditions for induced subgraphs of a PCG
to remain PCGs, and show how to reduce a PCG to smaller PCGs or
construct a larger PCG (resp., non-PCG) from a given PCG (resp., non-PCG).
For this, we first review the case when an induced subgraph of a PCG $G$
is a connected component of the graph.

\medskip
\noindent {\bf Components.}
It is known that a graph is a PCG if and only if each connected component of it is a PCG.
 The only if part trivially follows from Lemma~\ref{le:basic}.
 The if part is also easy to see:
 choose an inner vertex from the PCT  of a PCR of each connected component of  $G$,
 where we assume that  $d_{max}=1$ for all PCRs,
 and join the inner vertices to a new vertex with an edge weighted by a positive real $>1$
 to get a single tree whose leaf set is $V(G)$.
We easily see that the resulting tree is  a PCT for a PCR to $G$, showing that $G$ is a PCG.
It would be natural to consider similar properties on 2-edge-connected components
(resp., biconnected components) of graphs with bridges (resp., cut-vertices).
In fact, we show that the above property also holds for biconnected components.

\begin{lemma}\label{le:bi-conn}
Let a graph $G$  consist  of biconnected components
$B_i$, $i=1,2,\ldots,p$.
Then $G$ is a PCG if and only if each biconnected component $B_i$ of $G$ is a PCG.
\end{lemma}
\noindent {\bf Proof.}
The only if part trivially follows from Lemma~\ref{le:basic}.
To show the if part, it suffices to consider the case where $G$
  consists of two PCG graphs $G_1$ and $G_2$
such that $|V(G_1)\cap V(G_2)|=1$.

\begin{figure}[htbp] \begin{center}
 \includegraphics[scale=0.9]{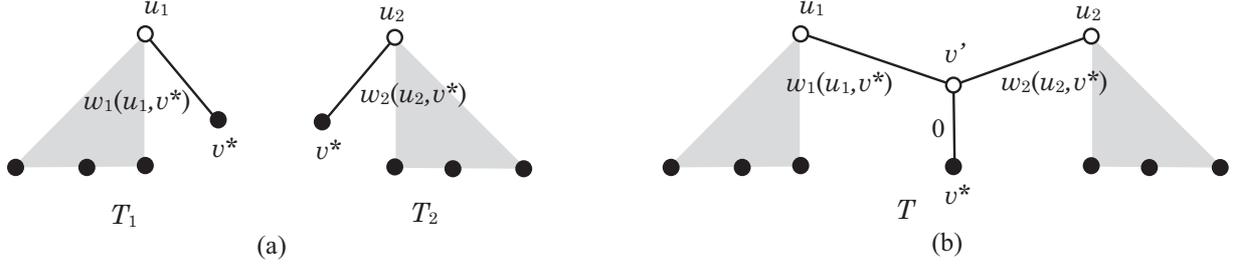} \end{center}
 \caption{(a) A normalized PCR  $(T_i, w_i, d_{\min}=\alpha,  d_{\max}=1)$ for each $i=1,2$,
 (b) The weighted tree $(T,w)$ obtained from $(T_i,w_i)$, $i=1,2$
 by joining edges $u_1v^*$ and $u_2v^*$ with a new inner vertex $v'$,
  where $w(u_1v')=w_1(u_1v^*)$,
 $w(u_2v')=w_2(u_2v^*)$ and $w(v'v^*)=0$
 } \label{f1:cutvertex}
 \end{figure}

Let $v^*\in V(G_1)\cap V(G_2)$.
By Lemma~\ref{le:2},
we see that, for a real $\alpha>0$,
 each PCG $G_i$ $(i=1,2)$ admits
 a normalized PCR  $(T_i, w_i, d_{\min}=\alpha,  d_{\max}=1)$,
 as illustrated in Figure~\ref{f1:cutvertex}(a).
Since they are normalized,  it holds that $w_i(e)>1/4$  for each  leaf edge  $e$  in $T_1$ and $T_2$.

Now we join the two PCRs by replacing
the leaf edge $u_iv^*$ in $T_i$ $(i=1,2)$ with a new inner vertex $v'$ and
three edges $u_1v'$, $u_2v'$ and $v'v^*$ setting their weights by
$w(u_1v^*):=w_1(u_1v^*)$,
$w(u_2v^*):=w_2(u_2v^*)$ and $w(v'v^*):=0$, respectively.
See Figure~\ref{f1:cutvertex}(b) for an illustration of the operation.
Let $(T,w)$ denote the resulting edge-weighted tree, and let $G'$ be  the graph $G(T,w,\alpha,1)$.
We will show that $G'$ is isomorphic to the graph $G$.

Since $w_i(e)>1/4$
     for each  leaf edge  $e$  in $T_i$ with $i=1$ and $2$,
     we see that    $  \mathrm{d}_{T,w}(u,v)> 4\cdot (1/4)=1=d_{\max}$
     for any pair of vertices $u\in L(T_1)-\{v^*\}$ and $v \in L(T_2)-\{v^*\}$.
This implies that $uv\not\in E(G')$.
Obviously for each $i=1,2$ and any pair $\{u,v\}\subseteq V(T_i)$,
it holds that $uv\in E(G')$ if and only if $uv\in E(G_i)$.
Therefore $G'$ is isomorphic to $G$, and $G$ is a PCG. \qed
\bigskip

Lemma~\ref{le:bi-conn} is a powerful tool to construct PCGs.
We can use it to `join' small PCGs into
a large PCG  to find  new subclasses of PCGs.
An edge or a single cycle has been shown to be  a PCG~\cite{Y-PCG}, and
a cactus is a graph with each biconnected component being
a cycle or an edge.
By simply applying Lemma~\ref{le:bi-conn}, we see the next.

\begin{lemma}\label{le:cactus}
Every cactus is a PCG.
\end{lemma}

A special case of cacti  (where each biconnected component is a cycle) was shown to be a subclass of PCG~\cite{Y-PCG}. However, by using Lemma~\ref{le:bi-conn}, we can greatly simplify the proofs~\cite{Y-PCG}.
Furthermore, Lemma~\ref{le:bi-conn} can be used to construct PCGs of   more complicated structures.


\medskip
\noindent {\bf Twins.}
Since twins have similar structures, we are interested to know wether PCG remains close under the operation of adding
a twin of a vertex.
This problem has been considered by Calamoneri et al.~\cite{exploring}.
They found that this property holds for false twins and
raised the case for true twins as an interesting open problem.
We will answer their question by exploring the property of true twins.

For false twins, the following lemma has been proven~\cite{exploring}.
We show that this can be proven by using normalized PCR.

\begin{lemma}\label{le:copy1}
Let $G$ be a graph with  false twins $v_1$ and $v_2$.
Then $G$ is a PCG if and only if $G-v_1$ is a PCG.
\end{lemma}
\noindent {\bf Proof.}
The only if part trivially follows from Lemma~\ref{le:basic}.
We show the if part assuming that  $G'=G-v_1$ is a PCG.
By Lemma~\ref{le:2}, we know that there is a normalized PCR
  $(T',w',\alpha>0,1)$  of $G'=G-v_1$.
We replace
the leaf edge $v' v_2$ in $T'$
with a new leaf $v_1$ and a new inner vertex $v''$ and
three edges $ v' v''$, $v'' v_2$ and $v'' v_1$,
setting their weights by
$w(v'v_2):= w'(v' v_2)$ and $w(v'' v_2):=w(v'' v_1):=0$.
Let $(T,w)$ denote the resulting edge-weighted tree.
Since $\mathrm{d}_{T,w}(v_1,v_2)=0<\alpha$,
$v_1v_2$ is not an edge in the graph $G(T,w,\alpha,1)$.
For any other leaf $v\in L(T)$, it holds
$\mathrm{d}_{T,w}(v,v_1)=\mathrm{d}_{T',w'}(v,v_2)$; and
for any leaves $u,v\in L(T)-\{v_1,v_2\}$, it holds
$\mathrm{d}_{T,w}(u,v)=\mathrm{d}_{T',w'}(u,v)$.
Therefore $(T,w,\alpha,1)$ is a PCR of $G$.
\qed
\bigskip

Lemma~\ref{le:copy1} can also be used to construct PCGs.
Based on Lemma~\ref{le:bi-conn}, we can construct large PCGs having cut-vertices.
By using Lemma~\ref{le:copy1}, we can increase the connectivity of PCGs. For example,
for each cut-vertex in a PCG, we can add a false twin of it  to the graph to get another PCG.
Lemma~\ref{le:copy1} also implies the following result.

\begin{lemma}\label{le:complete}
Every complete $k$-partite graph is a PCG.
\end{lemma}

Note that for a complete $k$-partite graph, if we iteratively delete a vertex in a pair of false twins
as long as  false twins exist, finally we will get a clique of $k$ vertices.
It is trivial that a clique is a PCG. By Lemma~\ref{le:copy1}, we know that
any complete $k$-partite graph is a PCG.
In fact, complete $k$-partite graphs contain many interesting graphs.
For examples, $\mathrm{K}_{1,2,2}$ is a $5$-wheel, $\mathrm{K}_{2,2,2}$ is an octahedron,
$\mathrm{K}_{1,2,4}$ is a $(4,3)$-fan, $\mathrm{K}_{2,2,5}$ is a $(4,5)$-cone,
$\mathrm{K}_{4,4,4}$ is a circulate graph $\mathrm{Ci}_{12}(1,2,4,5)$, and so on.
Some of them have been shown to be PCGs by using different techniques in the literature.

\medskip


Next, we consider true twins.
In fact, the statement in Lemma~\ref{le:copy1}   for true twins is no longer correct
because there is an example of a non-PCG $G$ such that   deleting a vertex in   true twins
 results in a  PCG.

\begin{figure}[htbp] \begin{center}
 \includegraphics[scale=1.0]{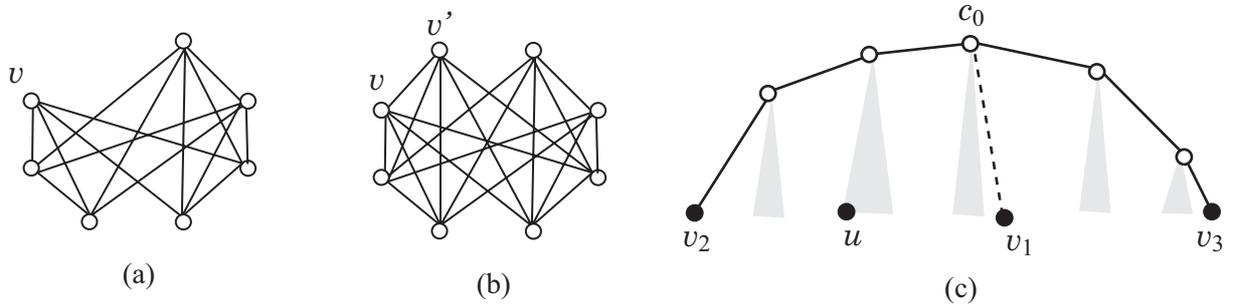} \end{center}
 \caption{(a) A graph $G$ with seven vertices, and
 (b) A graph $G'$ obtained from $G$ by adding a vertex $v'$ so that $v$ and $v'$ are true twins
 in $G'$,
 (c) A PCT $T$ obtained from $T'$ with a new leaf edge $c_0v_1$
 } \label{f1:example2}
 \end{figure}

The graph $G$ in Figure~\ref{f1:example2}(a) has only seven vertices.
This is a PCG since it has been proved that any
graph with at most seven vertices is a PCG~\cite{CFS2013}.
The graph $G'$  in Figure~\ref{f1:example2}(b) is obtained from the graph $G$
by a copy $v'$ of vertex $v$ so that $v$ and $v'$ form  true twins in $G'$.
The graph $G'$ has been shown to be a non-PCG~\cite{noPCG}.

We show that a non-PCG remains to be a non-PCG
after removing one of three  true twins.

\begin{lemma}\label{le:adjacent}
Let $G$ be a graph with three true twins $v_1,v_2$ and $v_3$, i.e., $N_G[v_1]=N_G[v_2]=N_G[v_3]$.
Then $G$ is a PCG if and only if $G-v_1$ is a PCG.
\end{lemma}
\noindent {\bf Proof.}
The only if part trivially follows from Lemma~\ref{le:basic}.
We show the if part assuming that  $G'=G-v_1$ is a PCG.
Let $(T',w',d'_{\min}, d'_{\max})$ be a   PCR of $G'$,
where $|V(T')|\geq |L(G)|\geq 2$.
We will construct a PCR $(T,w,d'_{\min}, d'_{\max})$ of $G$.


Let $c_0$ be the middle point of the path between $v_2$ and $v_3$ in $T'$,
i.e., $c_0$ is an inner vertex or an interior point on an edge
such that $d_{T',w'}(v_2,c_0)=d_{T',w'}(c_0,v_3)$.

We add $v_1$ to $T'$ as a new leaf creating a new edge between $v_1$ and $c_0$ in $T'$
 to construct a tree $T$ with $L(T)=V(G)$.
We set the edge weight $w(v_1c_0):={1 \over 2} d_{T',w'}(v_2,v_3)$.
If $c_0$ is an interior point on an edge $u_1u_2$ in $T'$, then we subdivide  $u_1u_2$
 into two edges $u_1c_0$ and $c_0u_2$ setting their weights so that
  $w(u_1c_0)+w(c_0u_2)=w'(u_1u_2)$ and
$c_0$ is still the middle point of the path between $v_2$ and $v_3$ in $T$.
For all other edges $e$ in $T'$, we set $w(e):=w'(e)$.
Note that $d_{T,w}(v_1,c_0)=d_{T,w}(v_2,c_0)=d_{T,w}(v_3,c_0)={1 \over 2}d_{T,w}(v_2,v_3)={1 \over 2}d_{T',w'}(v_2,v_3)$.
To prove that $(T,w, d'_{\min},  d'_{\max})$ is a PCR of $G$,
it suffices to prove that
for each vertex $u\in V(G)\setminus\{v_1\}$,
 $d_{T,w}(v_1,u)$ is equal to $d_{T,w}(v_i,u)$ for $i=2$ or $3$,
 which implies that
 $v_1u\in E(G)$ if and only if $v_iu\in E(G')$.
 Recall that $v_2u\in E(G')$ if and only if $v_3u\in E(G')$
 by assumption of $N_G[v_2]=N_G[v_3]$.

Let $u\in V(G)\setminus\{v_1\}$, where we assume without loss of generality
that $d_{T,w}(v_2,u)\leq d_{T,w}(v_3,u)$,
which means that the path between $u$ and $v_3$ passes through $c_0$ in $T'$,
 as illustrated in Figure~\ref{f1:cutvertex}(c).
Hence $d_{T,w}(v_3,u)=d_{T,w}(v_1,u)$ holds, as required.
\qed
\bigskip

Lemma~\ref{le:adjacent} implies that
a PCG with  true twins $u_1$ and $u_2$ can be augment to a larger PCG
with any number of new vertices $u_2,\ldots,u_k$ so that
every two vertices $u_i$ and $u_j$, $1\leq i<j\leq k$ form true twins.

\section{Conclusions}\label{sec:conclude}

In this paper, we have introduced some reduction rules on PCGs.
By using these rules, we can find more subclasses of PCG and
 simplify some arguments in previous papers.
Also the reduction rules can be used to find a class of non-PCGs
by   constructing lager non-PCGs from a given non-PCG in a similar way.
All graphs with at most seven vertices are known to be PCGs,
and a non-PCG with eight vertices has been found.
To find all non-PCGs with  $n=8$ vertices, the reduction rules can be used
to eliminate graphs with false twins or cut-vertices
 from the class of simple graphs with $n=8$ vertices,
 because such graphs are reduced to graphs with at most seven vertics which are all PCGs.
It is interesting to find more reduction rules on PCG.


\end{document}